\documentclass[aps,prd,11pt,floatfix]{revtex4}
\usepackage{graphicx}
\usepackage{color}
\newcommand{\half}{\frac{1}{2}}
\newcommand{\tr}{\mbox{tr}}

\newcommand{\be}{\begin{equation}}
\newcommand{\ee}{\end{equation}}
\newcommand{\bea}{\begin{eqnarray}}
\newcommand{\eea}{\end{eqnarray}}
\newcommand{\nn}{\nonumber}
\begin{document}
\begin{center}
{\large \bf  Numerical Investigation of Monopole Chains}
\end{center}
\centerline{ Gerald V.~Dunne and Vishesh Khemani }
\begin{center}
Department of Physics\\
University of Connecticut\\
Storrs, CT 06269-3046
\end{center}
\smallskip
\centerline{\large\bf Abstract}
\bigskip
{\small 
We present numerical results for chains of SU(2) BPS monopoles constructed from Nahm data. The long chain limit reveals an asymmetric  behavior transverse to the periodic direction, with the asymmetry becoming more pronounced at shorter separations. This analysis is motivated by a search for semiclassical finite temperature instantons in the 3D SU(2) Georgi-Glashow model, but it appears that in the periodic limit the instanton chains either have logarithmically divergent action or wash themselves out.
}

\bigskip

\leftline{\it \small Keywords:~\parbox[t]{15cm}{
monopole, instanton, finite temperature}}

\leftline{\it \small PACS:~\parbox[t]{15cm}{11.10.Kk, 11.15.Kc}}
%
%\newpage
\bigskip

In this note we investigate numerically the properties of long chains of $SU(2)$ BPS monopoles. In addition to the inherent mathematical interest in a search for such solutions, our physical motivation is to seek caloron (finite temperature instanton) solutions in the 3D $SU(2)$ Georgi-Glashow (GG) model. At $T=0$, such 3D GG instantons are mathematically identical to the 't Hooft-Polyakov monopoles \cite{thooft,polyakov,manton} of a $3+1$ dimensional Yang-Mills-Higgs theory; and at $T>0$, calorons of the 3D GG model would be periodic in the Euclidean time direction, with period $\beta=\frac{1}{T}$, and so would be "periodic monopoles". Recall that in 4D Yang-Mills theory, explicit finite action caloron solutions are known, both with trivial \cite{harrington} and nontrivial \cite{leeyi,lee,kraan} holonomy along the Euclidean time direction. But in the 3D GG model, no finite action caloron solutions are known. This lack of caloron solutions is somewhat surprising for the following reason. At $T=0$ the 3D GG model is confining due to nonperturbative instanton effects \cite{polyakovgg}. More recently it has been shown analytically \cite{dkkt}, and confirmed by a lattice analysis \cite{barresi}, that the 3D $SU(2)$ GG model has a deconfining phase transition in the ${\bf Z}_2$ universality class at nonzero $T$, and that the critical behavior at this transition involves both instanton and charged scalar degrees of freedom. It is therefore somewhat surprising that it has not been possible to construct finite action semiclassical solutions corresponding to finite $T$ instantons. 

To establish notation, consider the $SU(2)$ Georgi-Glashow (GG) model in 3 dimensional spacetime. This model is an $SU(2)$ Yang-Mills gauge theory minimally coupled to a scalar field, $\Phi$, in the adjoint representation, and with a symmetry breaking quartic scalar potential. The Euclidean action is
\be
S = \int d^3 x \left[ \half \tr\left( F_{\mu\nu} F_{\mu\nu} \right) + \half D_\mu \Phi^a D_\mu \Phi^a + \frac{\lambda}{4}\left( \Phi^a \Phi^a - v^2 \right)^2  \right] \, .
\label{action}
\ee
Here $A_\mu = \half A_\mu^a \tau^a$, $
F_{\mu\nu}= \partial_\mu A_\nu - \partial_\nu A_\mu - i g [A_\mu , A_\nu]$, 
$\Phi =\half \Phi^a \tau^a$, and $D_\mu \Phi = \partial_\mu \Phi - i g [A_\mu , \Phi]$, where  $\tau^a$ ($a=1, 2, 3$) denotes the $2\times 2$ Pauli matrix generators of $su(2)$. Note that in $3$ dimensional spacetime the coupling $g$ has dimensions of $({\rm mass})^{1/2}$, as do the fields $\Phi$ and $A_\mu$. 3D GG instantons ("monopoles") are solutions to the classical Euclidean equations of motion:
\bea
D_\mu D_\mu \Phi &=& \lambda \left( \Phi^a \Phi^a - v^2 \right) \Phi \, , \nonumber \\
D_\mu F_{\mu\nu} &=& i g [D_\nu \Phi , \Phi] \, .
\label{eom}
\eea
No explicit periodic solutions to the instanton ("monopole") equations (\ref{eom})  are known. The most direct way to study multi-monopoles is by ansatz, either radial or axial. The axial ansatz can produce chains of alternating monopoles and antimonopoles \cite{kleihaus}, but does not appear to support chains of monopoles.  Instead, here we restrict ourselves to the BPS limit (the scalar self-coupling $\lambda\to 0$), where there exist other approaches to constructing monopoles. A hint of periodic monopoles appears in \cite{ford}, where doubly periodic 4D Yang-Mills instantons were studied -- as one period shrinks one is left with 3D BPS monopoles periodic in the other direction. But the truly periodic limit is not fully understood. Direct analyses \cite{cherkis,ward} have found periodic BPS monopoles, but with divergent energy. Also, Bielawski \cite{bielawski} has studied the n-monople problem for widely separated monopoles, showing that the moduli space is exponentially close to the hyperkahler Gibbons-Manton metric \cite{gibbons}.

In the BPS limit the instanton/monopole equations (\ref{eom}) simplify to
\bea
D_\mu \Phi=\frac{1}{2} \epsilon_{\mu\nu\sigma} F_{\nu\sigma}
\label{bps}
\eea
In the BPS limit the energy/action reduces to
\bea
S_{\rm BPS}=\int d^3 x \, \nabla^2 \, \tr\left(\Phi^2\right)
\label{bpsaction}
\eea
Thus, a convenient gauge invariant quantity to study is 
\bea
|\Phi|^2\equiv \frac{1}{2}\tr\left(\Phi^2\right).
\label{higgs}
\eea
In this paper we consider the Nahm construction \cite{nahm,manton} of BPS monopoles, in which one can reconstruct the monopole fields $\Phi$ and $A_\mu$ from a solution of the associated Nahm equations, as described below. 

\begin{figure}[ht]
\includegraphics[scale=0.5]{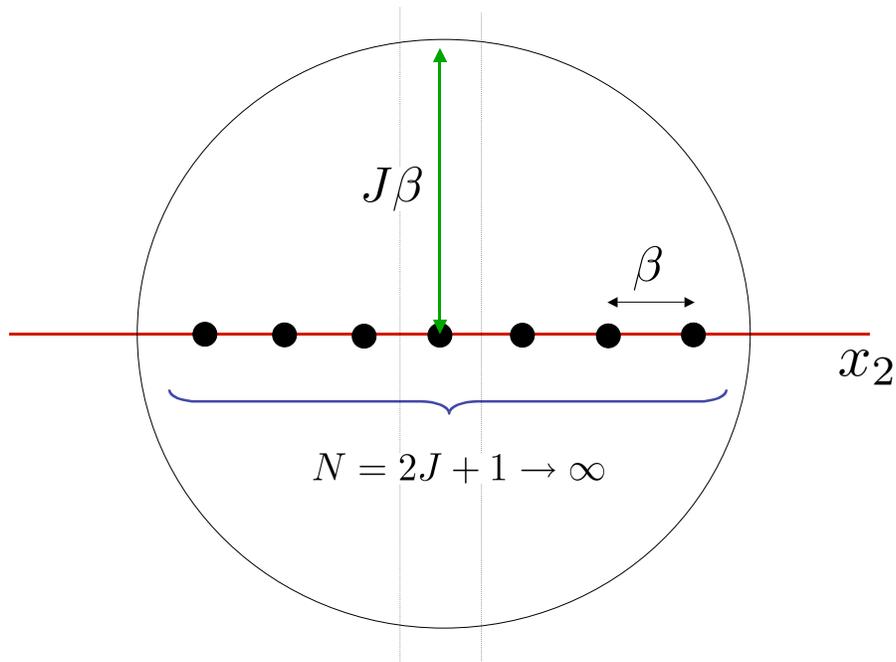}
\caption{An array of Dirac monopoles aligned along the $x_2$ axis with separation $\beta$. In the large $J$ limit, within the central slab region $-\frac{\beta}{2}\leq x_2\leq \frac{\beta}{2}$, the magnetic potential goes over from the 3D $1/r$ form to a 2D $\ln (\rho)$ form within the region of radius of order $J \beta$.}
\label{figure1}
\end{figure}

It is useful to consider what one might expect to find. Outside a core region (whose scale we take to be 1 in this paper) a single BPS monopole behaves like an abelian Dirac monopole. Thus, one might expect (at least when the period is greater than the core size, which corresponds to low $T$ in the 3D GG language) that a periodic BPS monopole would look like a periodic array of Dirac monopoles, outside of some core region around each monopole (see Figure 1). But if this were the case, the action/energy would be logarithmically divergent \cite{cherkis,ward}. This is because for an array of $N=2J+1$ Dirac monopoles, along the $x_2$ axis with separation $\beta$, the magnetic field is $\vec{B}=-\vec{\nabla}\phi$, where the magnetic scalar potential is
\bea
\phi=\sum_{n=-J}^{+J} \frac{1}{\sqrt{x_1^2+x_3^2+(x_2-n \beta)^2}}
\label{array}
\eea
For any finite $N$, at truly asymptotic distances $|\vec{x}|\gg J \beta$, the scalar potential $\phi$ behaves as $\phi\sim \frac{N}{|\vec{x}|}$, and the solution looks like a charge $N$ Dirac monopole. But at intermediate distances, inside a sphere of radius of order $J \beta$ (see Figure \ref{figure1}) the behavior is quite different. For a periodic array in the $J\to\infty$ limit we can restrict our attention to the central slab with $-\frac{\beta}{2}\leq x_2 \leq \frac{\beta}{2}$, and write \cite{cherkis}:
\bea
\phi={\rm constant}+\sum_{n=-\infty}^{\infty\,\,\prime} \left(\frac{1}{\sqrt{x_1^2+x_3^2+(x_2-n \beta)^2}}-\frac{1}{n \beta}\right)
\label{reg-array}
\eea
Here the sum is extended to infinity and the divergence at the origin has been subtracted (the prime indicates that the $n=0$ term is excluded in the sum).  
Within this slab, at transverse distances inside some region of order  $J \beta$, $\phi$ behaves as : $\phi\sim \ln \left(\frac{\rho}{2J+1}\right)$, where $\rho$ is the radial transverse distance: $\rho=\sqrt{x_1^2+x_3^2}$. This illustrates the physical origin of the transition from the 3D Laplace Green's function to the logarithmic 2D Laplace Green's function. So, assuming the periodic BPS monopole is like an array of Dirac monopoles outside a core region, we might expect that the Higgs field magnitude $|\Phi|$ would behave as $|\Phi|\sim \ln \rho$ at large $\rho$. This leads to logarithmically divergent action/energy, as can be seen from (\ref{bpsaction}) \cite{cherkis,ward}.

Is it possible to avoid this divergence? One possible solution is that the expectation of independent core regions is not realized. Since the BPS action/energy is proportional to the instanton/monopole charge, it is natural to expect that within the central slab region $-\frac{\beta}{2}\leq x_2 \leq \frac{\beta}{2}$ the net charge should be 1 even in the $J\to \infty$ limit,  and so it might be possible to define a large $J$ limit in which the action/energy stays finite. This could happen if the core region spreads out  significantly in the transverse directions. Our strategy here is simply to investigate numerically the large $J$ limit of a class of monopoles which are manifestly BPS solutions for any $J$ and which  correspond to $N=2J+1$ monopoles equally spaced along the $x_2$ axis.

We use the Nahm construction of monopoles \cite{nahm,manton}, which is based on first solving the "Nahm equations", a set of coupled ordinary differential equations [rather than the coupled partial differential equations in (\ref{bps})] :
\bea
\frac{d T_i}{ds}=\frac{1}{2}\,\epsilon_{ijk}[T_j, T_k ]
\label{nahmeqs}
\eea
Here $T_i(s)$, for $i=1, 2, 3$, and $s\in [-1,1]$, are $N\times N$ square matrices for an $N$-monopole, satisfying three conditions: (i) $T_i(s)$ is regular for $s\in (-1,1)$, and has simple poles at $s=\pm1$, with matrix residues defining irreducible representations of $su(2)$; (ii) $T^\dagger_i=-T_i$; and (iii) $T_i(-s)=T^T_i(s)$ (in some basis). It is not trivial to find solutions to (\ref{nahmeqs}) satisfying these conditions. Nevertheless, an explicit solution is known which corresponds to a chain of $N$ BPS monopoles equally spaced along a line \cite{ercolani,sutcliffe} (this generalizes an explicit 2-monopole solution in \cite{brown}):
\bea
T_1(s)&=&-\frac{{\bf K}}{2} {\rm ds}\left((s+1){\bf K} {\bigg |} k\right)\, R_1\nn\\
T_2(s)&=&-\frac{{\bf K}}{2} {\rm ns}\left((s+1){\bf K} {\bigg |} k\right)\, R_2\nn\\
T_3(s)&=&-\frac{{\bf K}}{2} {\rm cs}\left( (s+1){\bf K}{\bigg |} k\right)\, R_3
\label{nahmdata}
\eea
Here ds, ns and cs are elliptic functions, with real elliptic parameter $0\leq k < 1$, and ${\bf K}(k)$ is the elliptic quarter period. The constant matrices $R_i$ form an $N\times N$ irreducible representation of $su(2)$. For definiteness, we choose $R_3=-2 i \, {\rm diag}\left(J, J-1,  \dots, -J+1, -J\right)$, and $R_{1,2}$ to be antihermitean matrices with nonzero entries: $(R_1)_{n,n+1}=(R_1)_{n+1,n}= i\sqrt{ n(2J-n+1)}$ and $(R_2)_{n,n+1}=-(R_2)_{n+1,n}= -\sqrt{n(2J-n+1)}$. Since we are primarily interested in the central monopole in a long chain, we take $N=2J+1$ to be odd.

Given such Nahm data, to construct the monopole fields $\Phi$ and $A_\mu$ which satisfy the BPS equation (\ref{bps}), one first solves the 1D Weyl equation
\bea
\left\{ {\bf 1}_{2N} \frac{d}{ds}-{\bf 1}_N \otimes \vec{x}\cdot\vec{\tau} +i T_j(s)\otimes \tau_j\right\}v=0\quad ,
\label{weyl}
\eea
where $\vec{x}\in {\bf R}^3$ parameterizes the solution. There are two independent solutions, $v_a(s; \vec{x})$, [for $a=1, 2$],  of (\ref{weyl}) that can be normalized as
\bea
\int_{-1}^1ds \, v_a^\dagger \, v_b =\delta_{ab} \quad .
\label{norm}
\eea
Then the $su(2)$ monopole fields are given in terms of the $v_a(s; \vec{x})$ by the following construction:
\bea
\Phi_{ab}(\vec{x}) &=&i \int_{-1}^1ds \,s\, v_a^\dagger (s; \vec{x}) \, v_b(s; \vec{x})
\label{phifield}\\
A_j^{ab}(\vec{x})&=&\int_{-1}^1ds \,v_a^\dagger(s; \vec{x}) \,\frac{\partial}{\partial x_j} v_b(s; \vec{x})
\label{afield}
\eea
An efficient numerical scheme for solving (\ref{weyl}) has been presented in \cite{hs} where symmetric $N=3$ and $N=4$ monopoles were studied from their known Nahm data. We have used a similar numerical scheme, programmed in matlab. The main complication is that the Nahm data, which appears in (\ref{weyl}),  has singularities at the end points. Indeed, near $s=\mp1$ there are simple poles:
\bea
-i T_j(s)\otimes \tau_j &\sim& \frac{i}{2} \frac{\left(R_1\otimes  \tau_1+R_2\otimes  \tau_2 +R_3\otimes  \tau_3\right)}{1+s}\qquad , \quad s\to -1\\
-i T_j(s)\otimes  \tau_j &\sim& \frac{i}{2} \frac{\left(R_1\otimes  \tau_1+R_2\otimes  \tau_2 -R_3\otimes  \tau_3\right)}{1-s}\qquad , \quad s\to +1
\label{poles}
\eea
These residues have eigenvalues $J$ and $-(J+1)$, with degeneracies $2J+2$ and $2J$, respectively. 
The negative eigenvalues correspond to singular behavior of the vector $v(s)$ at the end-points and so must be suppressed in order to satisfy the normalization conditions (\ref{norm}). That is, as $s\to \mp 1$, $v$ must lie in the $2J+2$ dimensional subspace spanned by the eigenvectors of the associated residue matrix with positive eigenvalues. These subspaces are independent of $\vec{x}$ and of the elliptic parameter $k$ (which determines the period $\beta$), and so may be constructed once and for all for a given monopole number. Following \cite{hs}, we solve (\ref{weyl}) by shooting from either end of the interval, and matching at $s=0$. The intersection of these shoots leaves just two independent solutions, $v_1$ and $v_2$, each of which is a $2N$-component column vector, depending parametrically on $\vec{x}$.
Given $v_1$ and $v_2$, we construct the magnitude squared of the Higgs field  using (\ref{phifield}). We focus on $|\Phi|^2$ since it is gauge invariant, its zeros indicate the locations of the monopoles, and because it determines the BPS  action/energy density in (\ref{bpsaction}).
We implemented this numerical algorithm for three different values of the elliptic parameter:  $k=0.99$, $k=0.5$ and $k=0.2$, which we found correspond to period $\beta=3.25$, $\beta=0.8$ and $\beta=0.3$, respectively, measured in units where the single-monopole core size is 1.

\begin{figure}[h]
\vskip -2cm
\includegraphics[scale=0.45,angle=270]{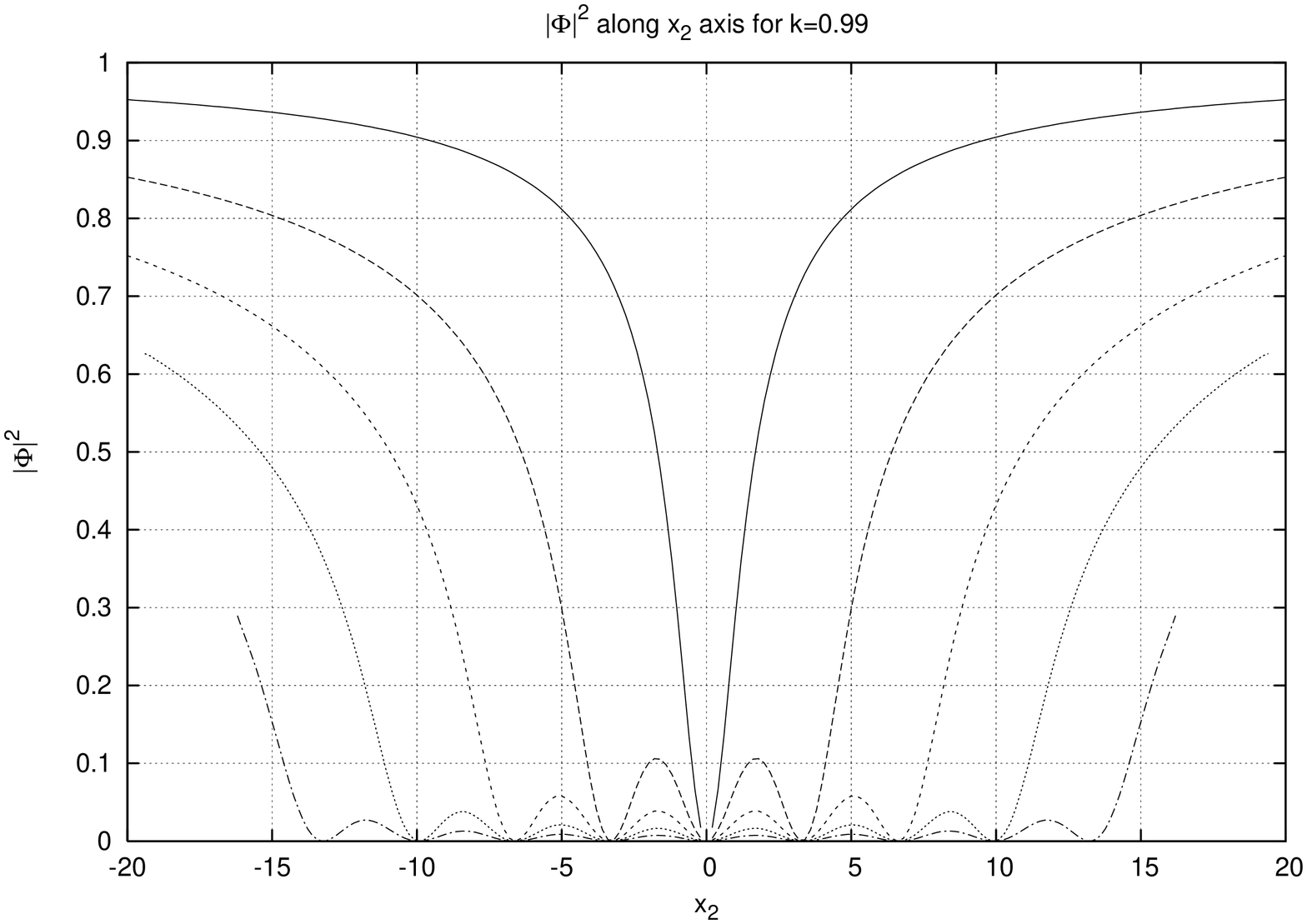}
\includegraphics[scale=0.45,angle=270]{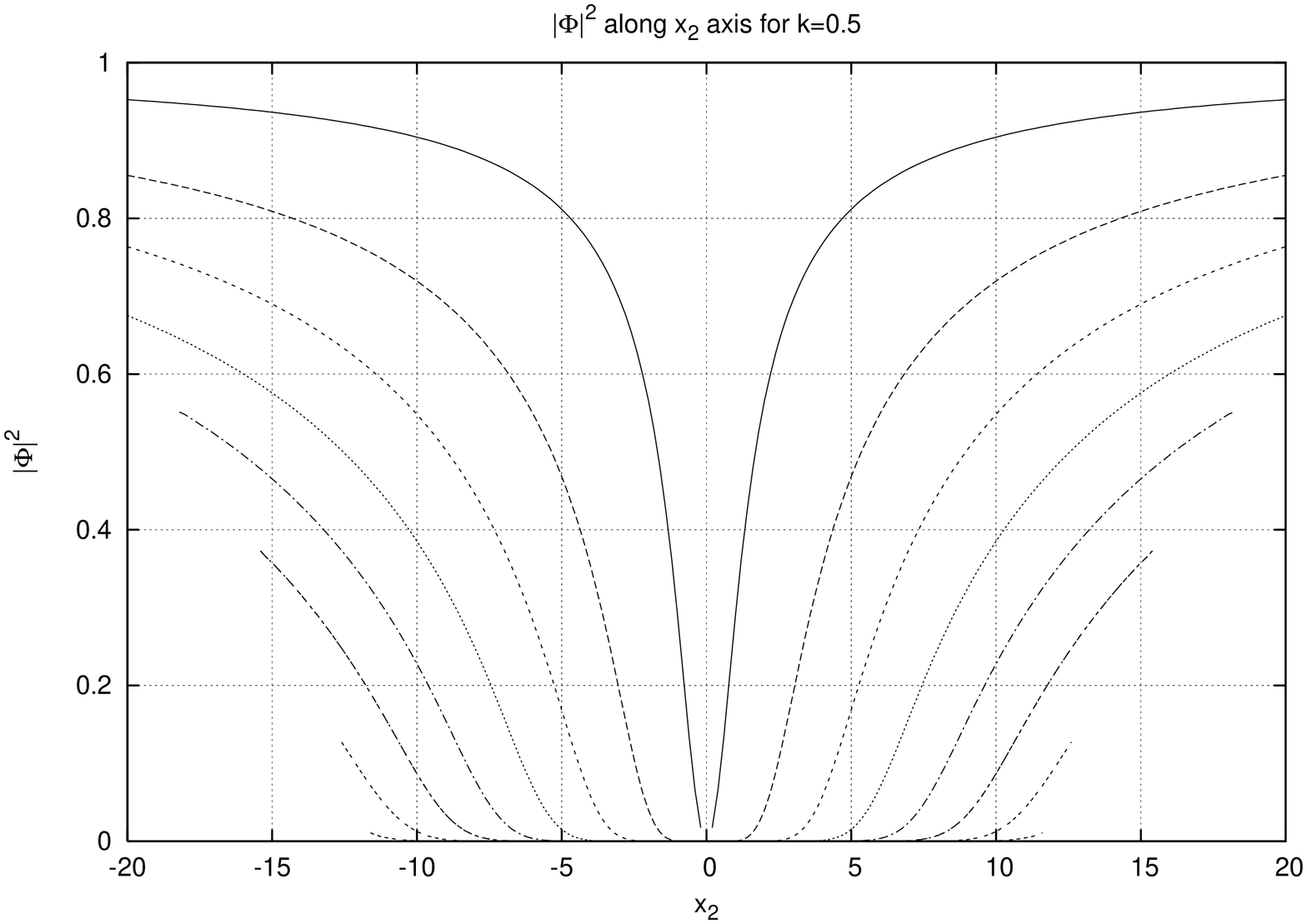}
\includegraphics[scale=0.45,angle=270]{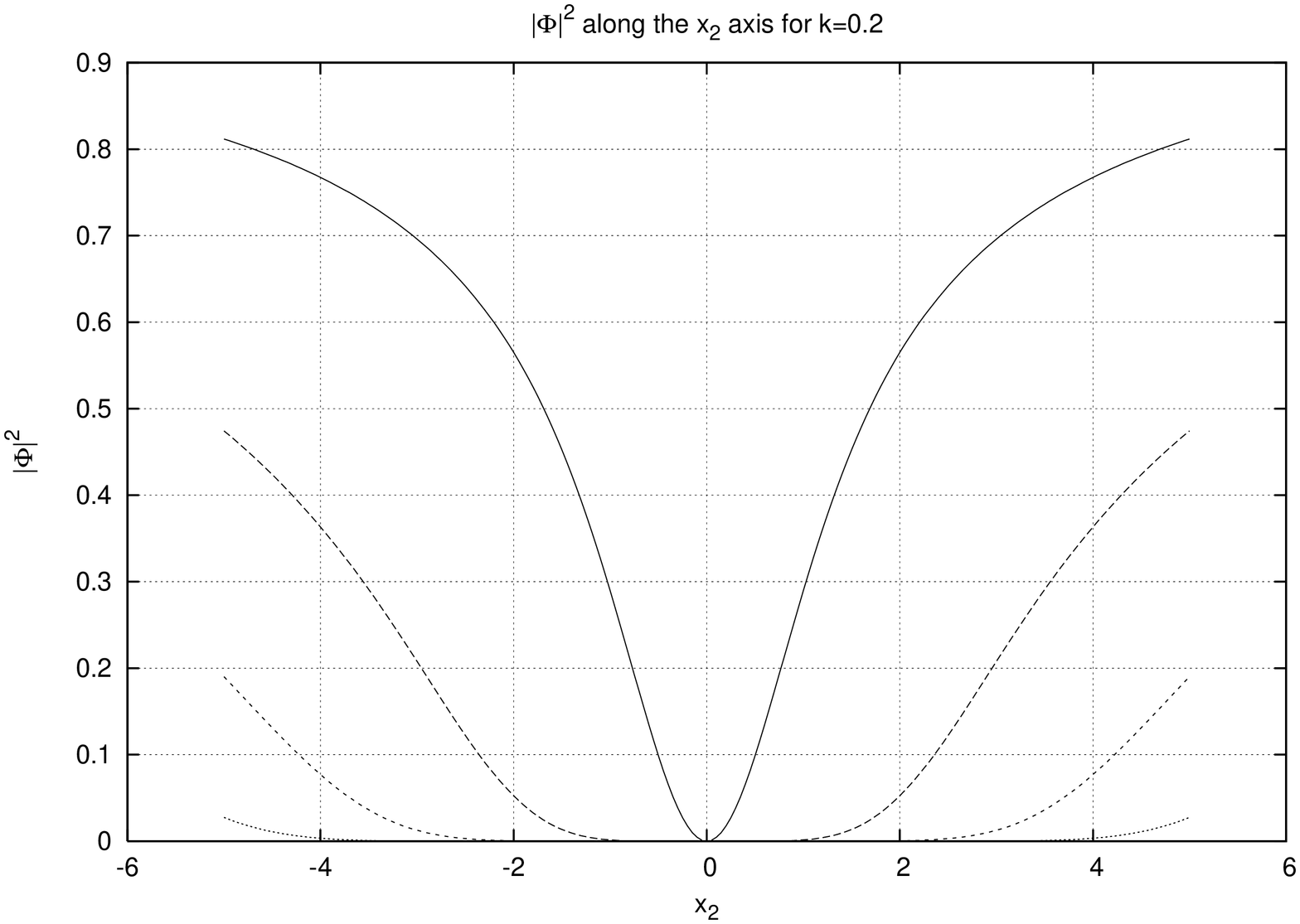}
\caption{Plots of $|\Phi|^2$ along the $x_2$ axis for elliptic parameter $k=0.99$ (top - Fig 2a), $k=0.5$ (middle - Fig 2b) and $k=0.2$ (bottom - Fig 2c). The different curves correspond to increasing values of $N$, starting with $N=1$ at the top. For $k=0.99$ one can see equally spaced zeros appearing along the $x_2$ axis,  but for $k=0.5$ and $k=0.2$ these zeros cannot be seen on this scale. The periodicity can be seen in Figure 3 which plots $\log(|\Phi|^2)$.}
\label{figure2}
\end{figure}

\begin{figure}[t]
\vskip -2cm
\includegraphics[scale=0.45,angle=270]{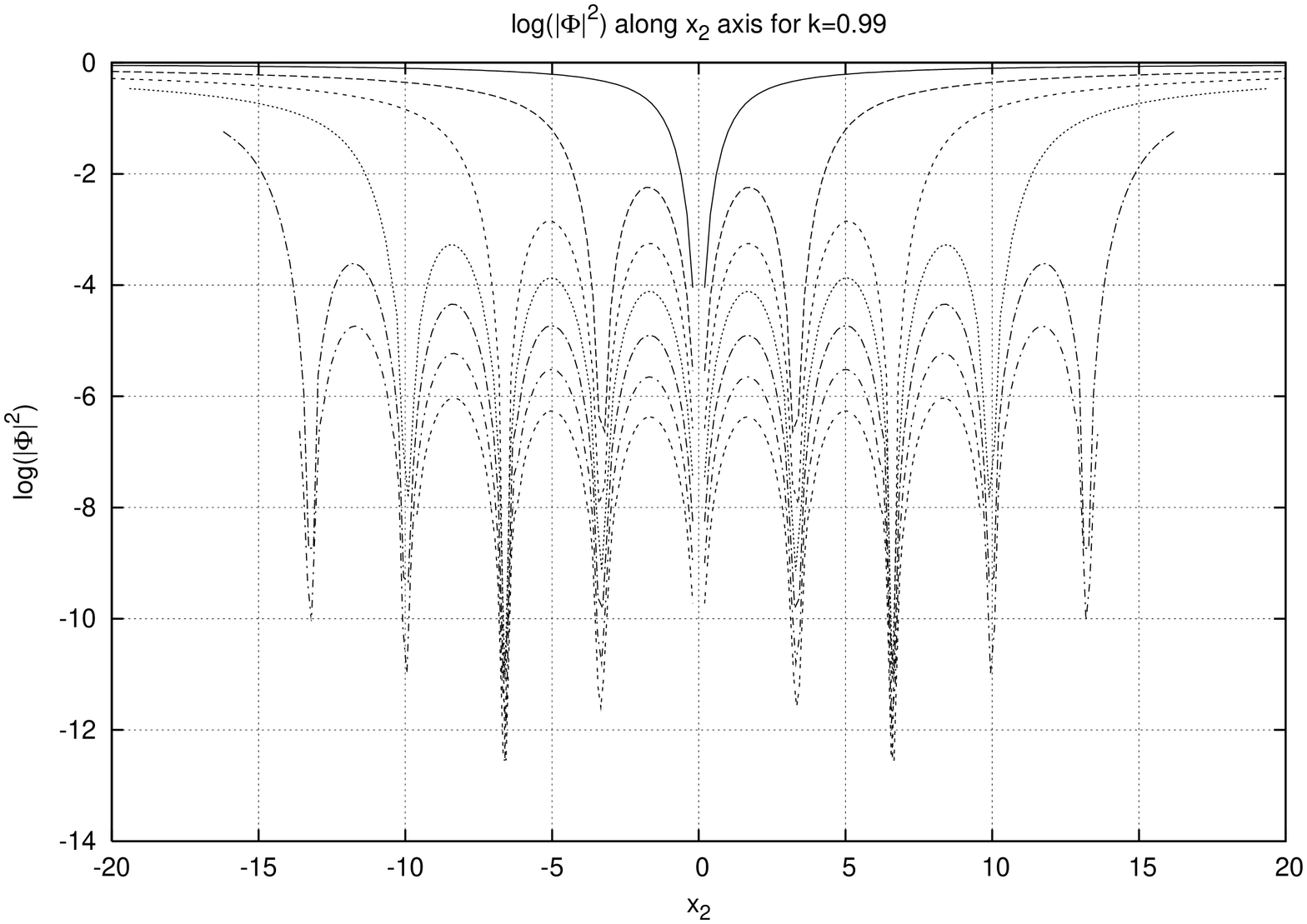}
\includegraphics[scale=0.45,angle=270]{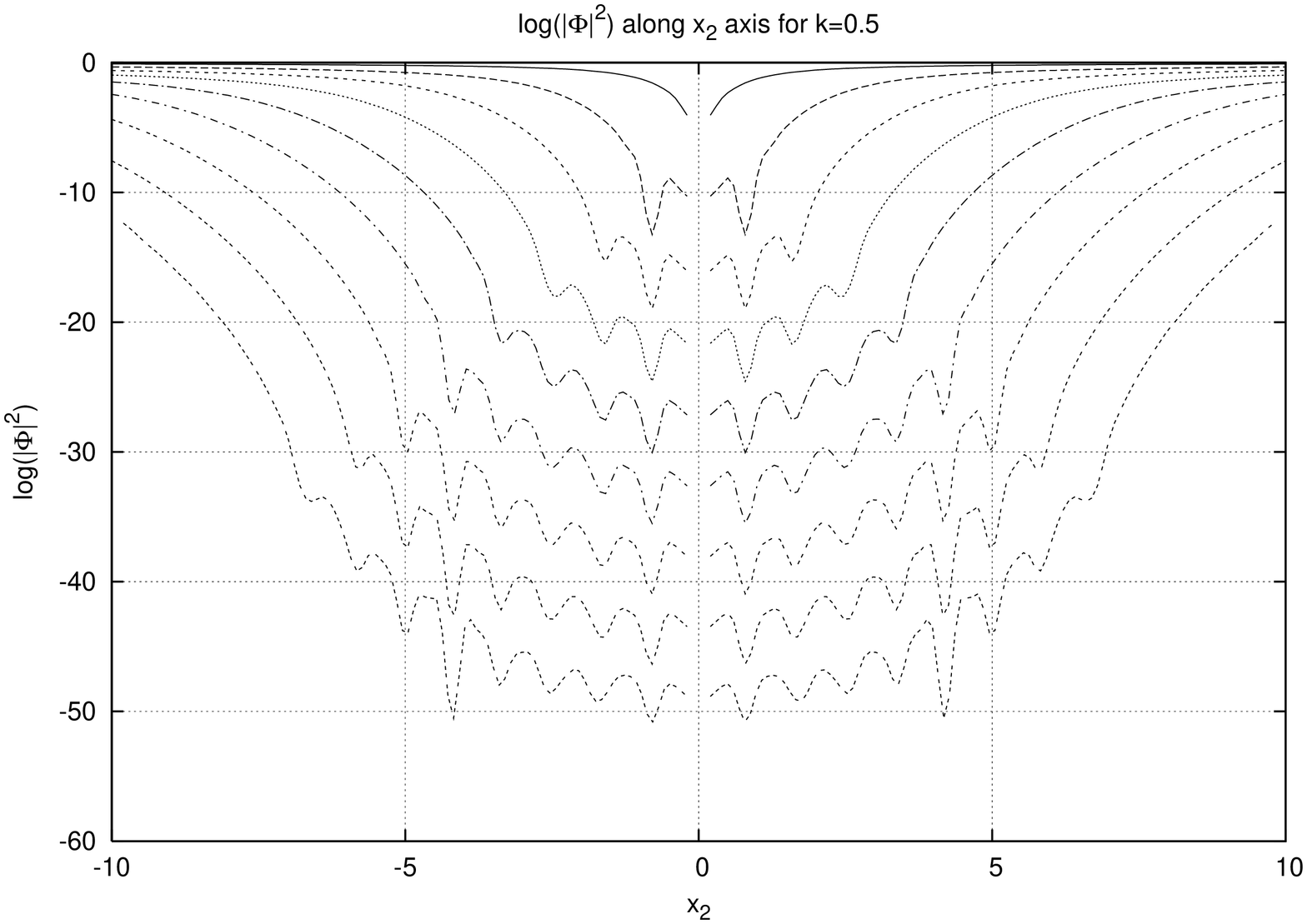}
\includegraphics[scale=0.45,angle=270]{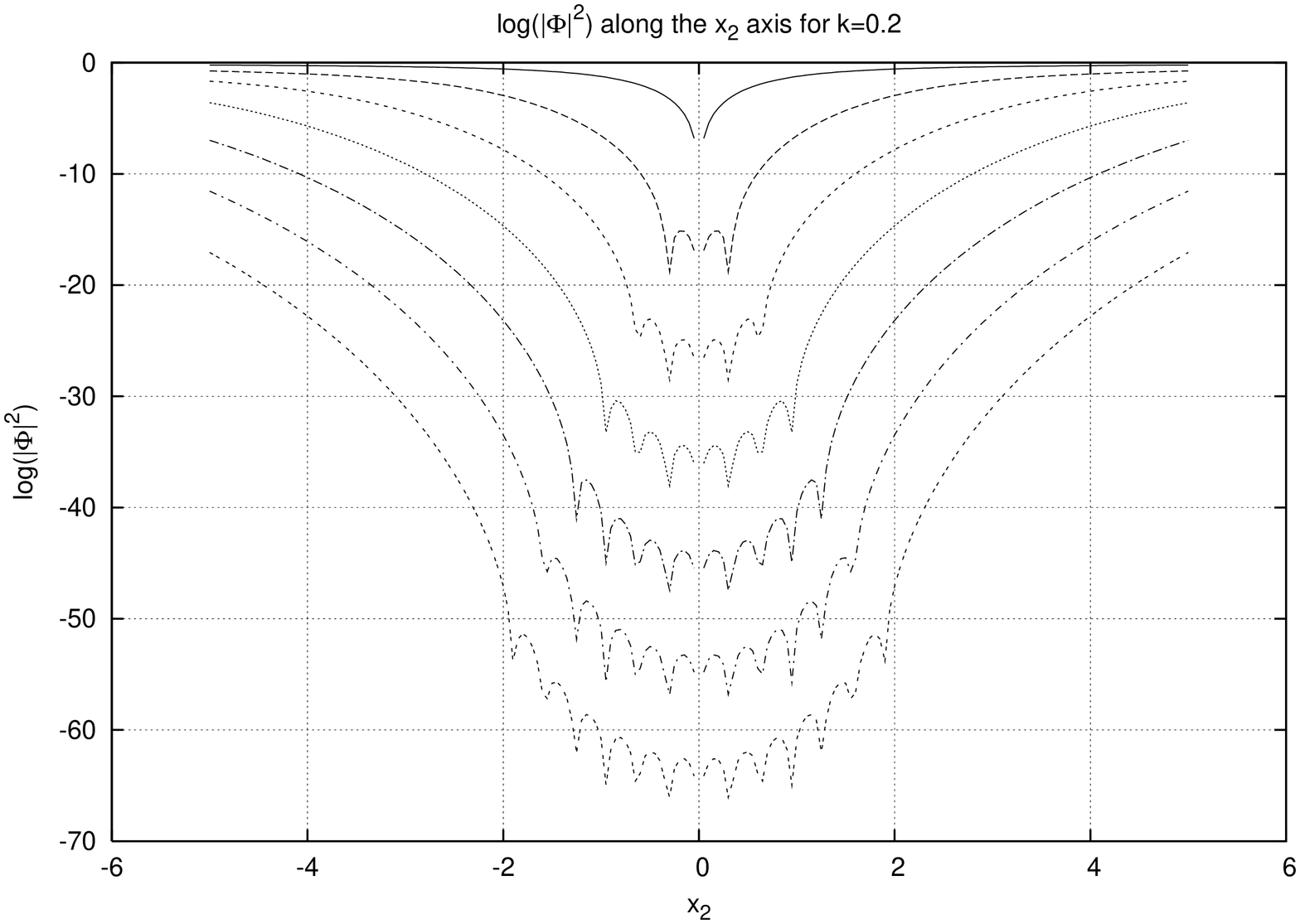}
\caption{Plots of  $\log(|\Phi|^2)$ along the $x_2$ axis for elliptic parameters $k=0.99$ (top - Fig 3a), $k=0.5$ (middle - Fig 3b) and $k=0.2$ (bottom - Fig 3c). The minima clearly indicate the periodic structure along the $x_2$ axis, and one can count the increasing number of monopoles in the chain for successive curves, beginning with $N=1$ at the top in each plot.}
\label{figure3}
\end{figure}

\begin{figure}[ht]
\vskip -2cm
\includegraphics[scale=0.45,angle=270]{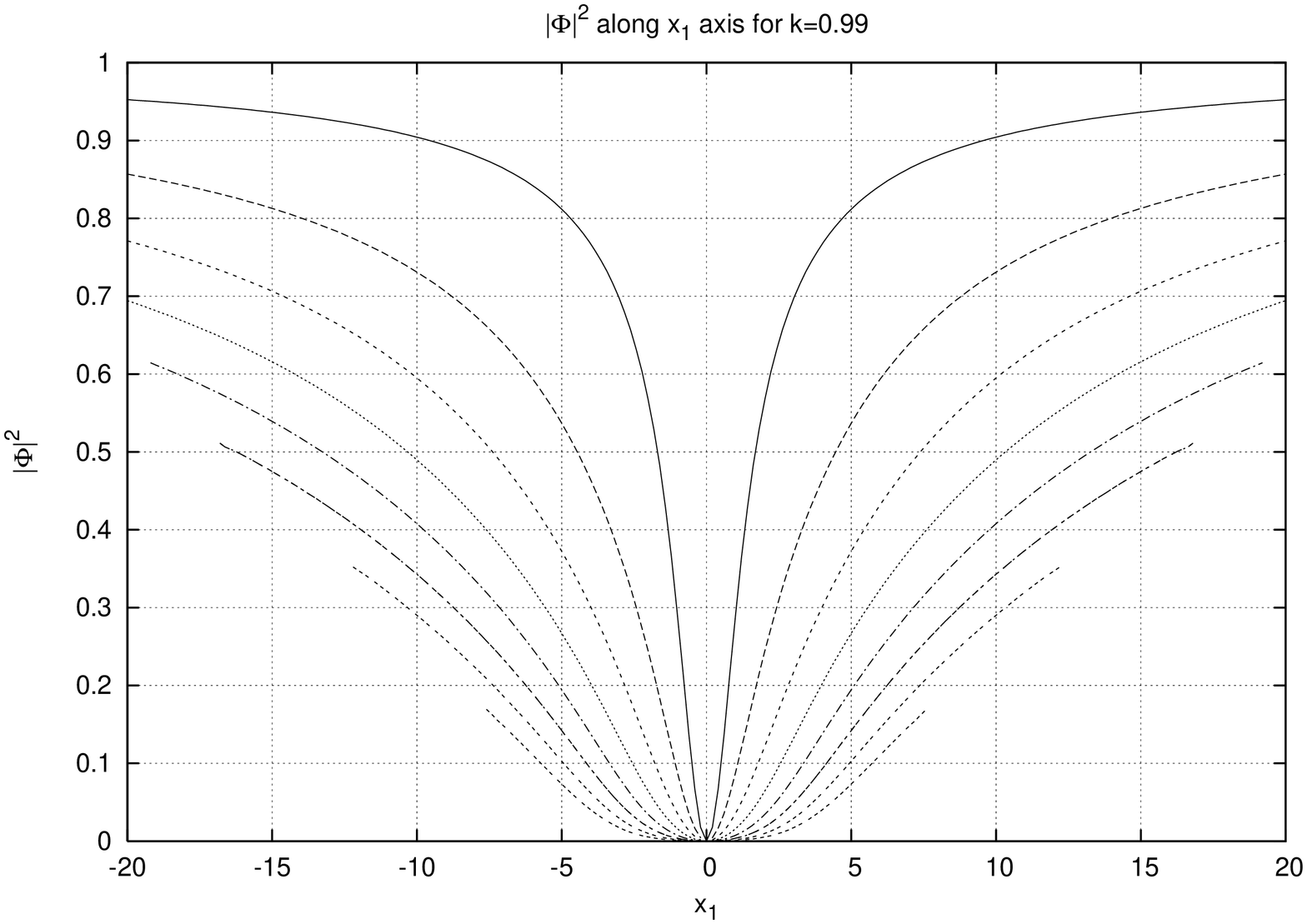}
\includegraphics[scale=0.45,angle=270]{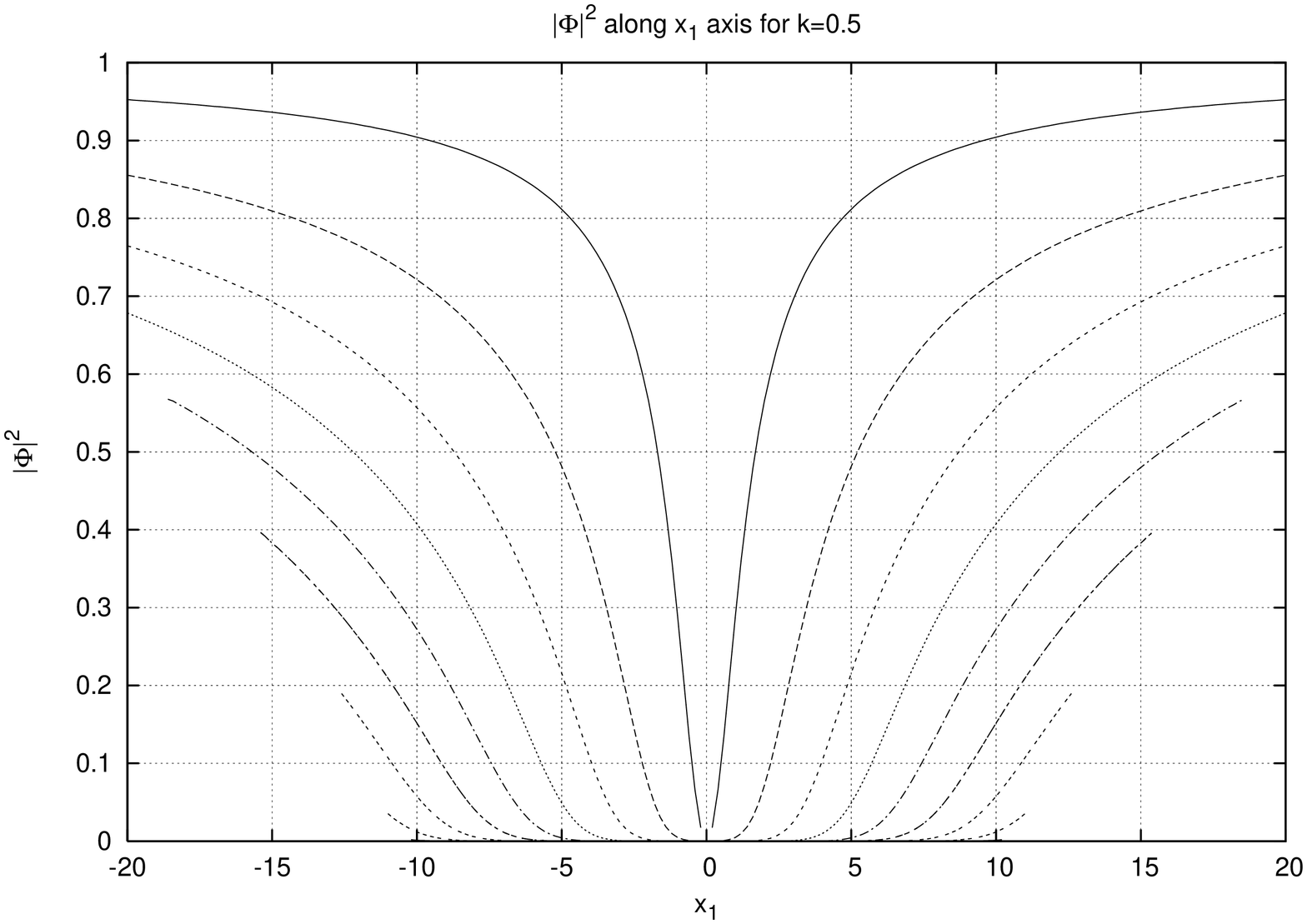}
\includegraphics[scale=0.45,angle=270]{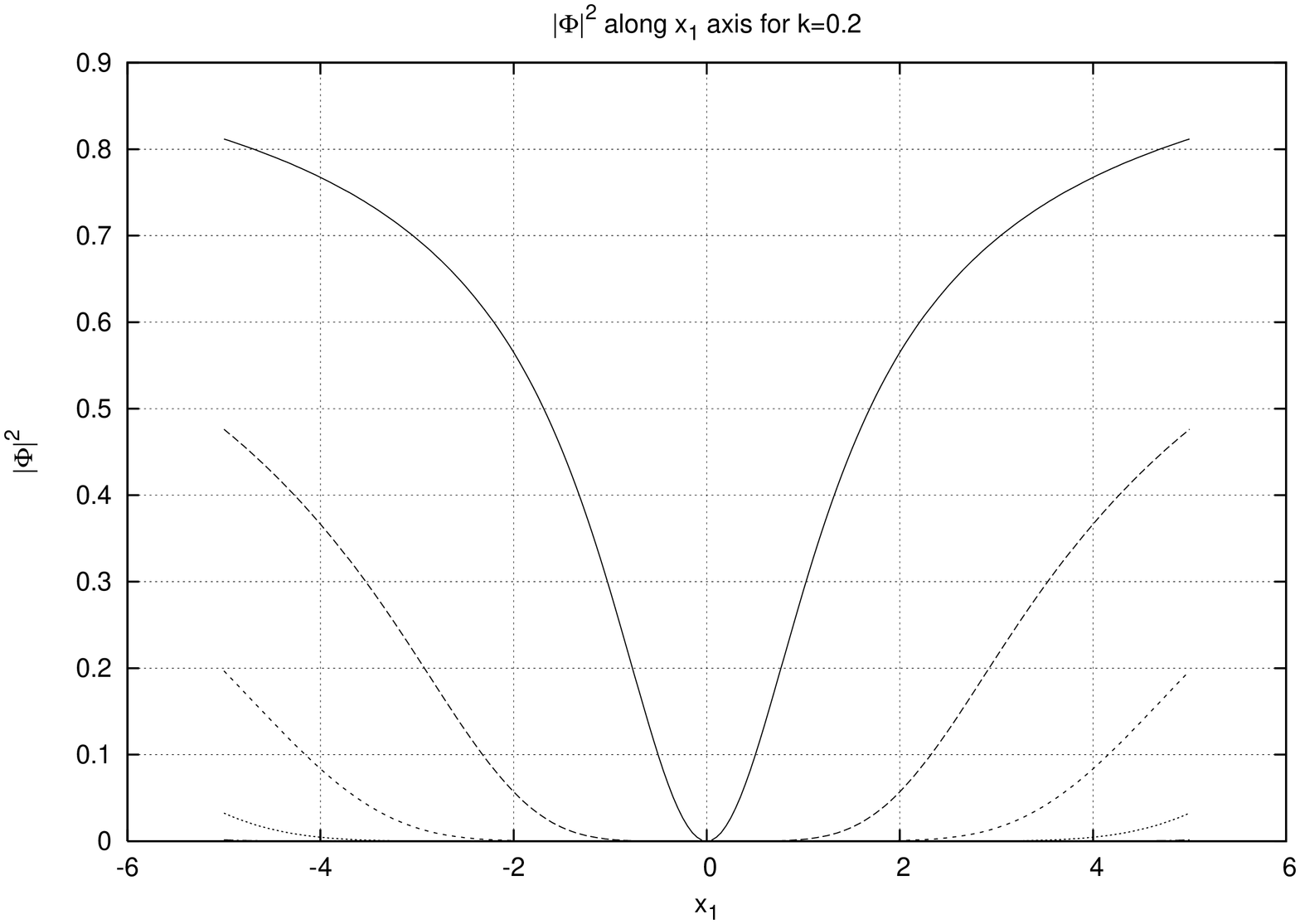}
\caption{Plots of $|\Phi|^2$ along the $x_1$ axis for elliptic parameter $k=0.99$ (top - Fig 4a), $k=0.5$ (middle - Fig 4b) and $k=0.2$ (bottom - Fig 4c). The different curves correspond to increasing values of $N$, starting with $N=1$ at the top. Note that for $k=0.5$ and $k=0.2$ these are very similar to the corresponding curves in Figure 2, but for $k=0.99$ there is a marked difference from the corresponding curve in Figure 2.}
\label{figure4}
\end{figure}

\begin{figure}[ht]
\vskip -2cm
\includegraphics[scale=0.45,angle=270]{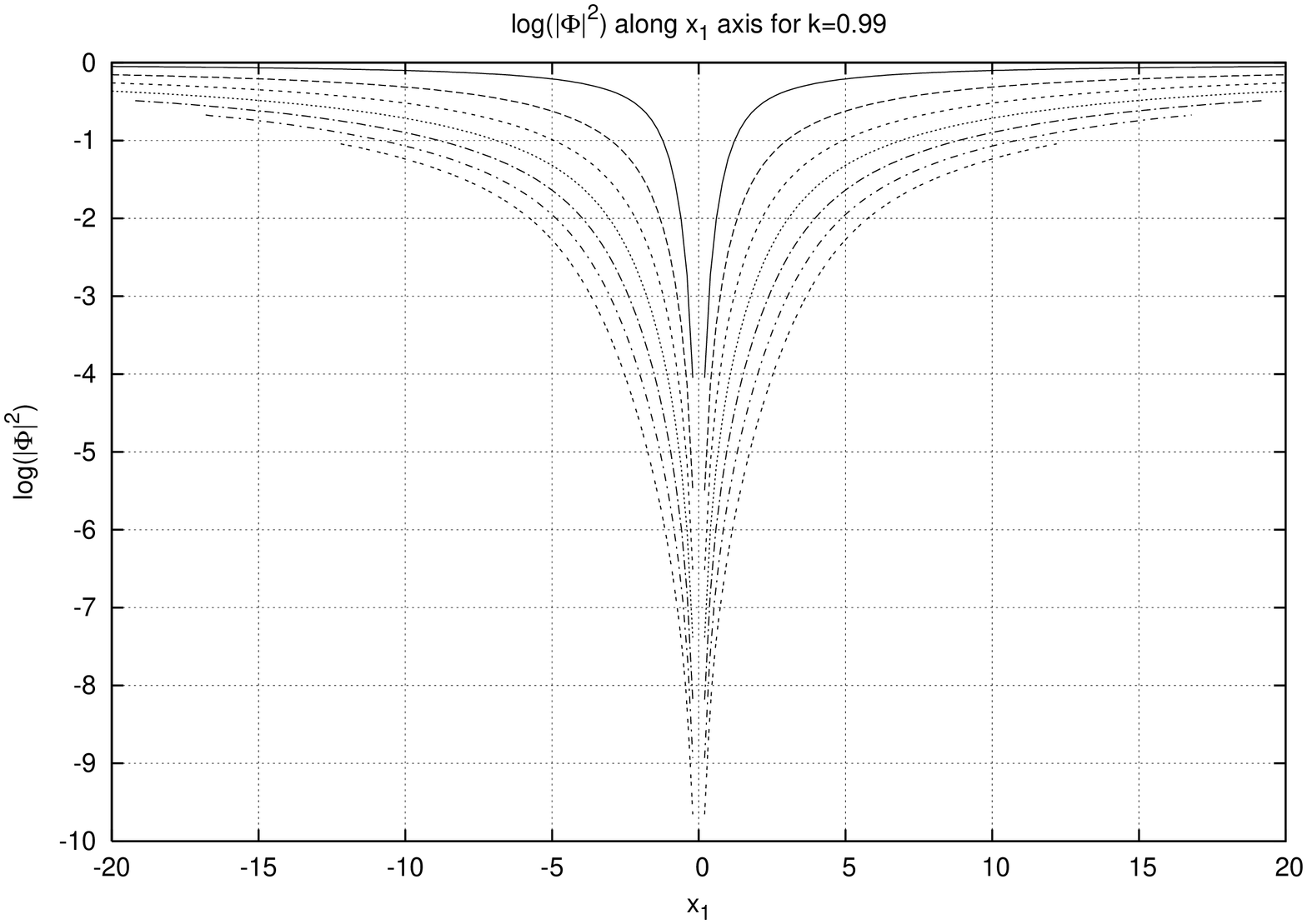}
\includegraphics[scale=0.45,angle=270]{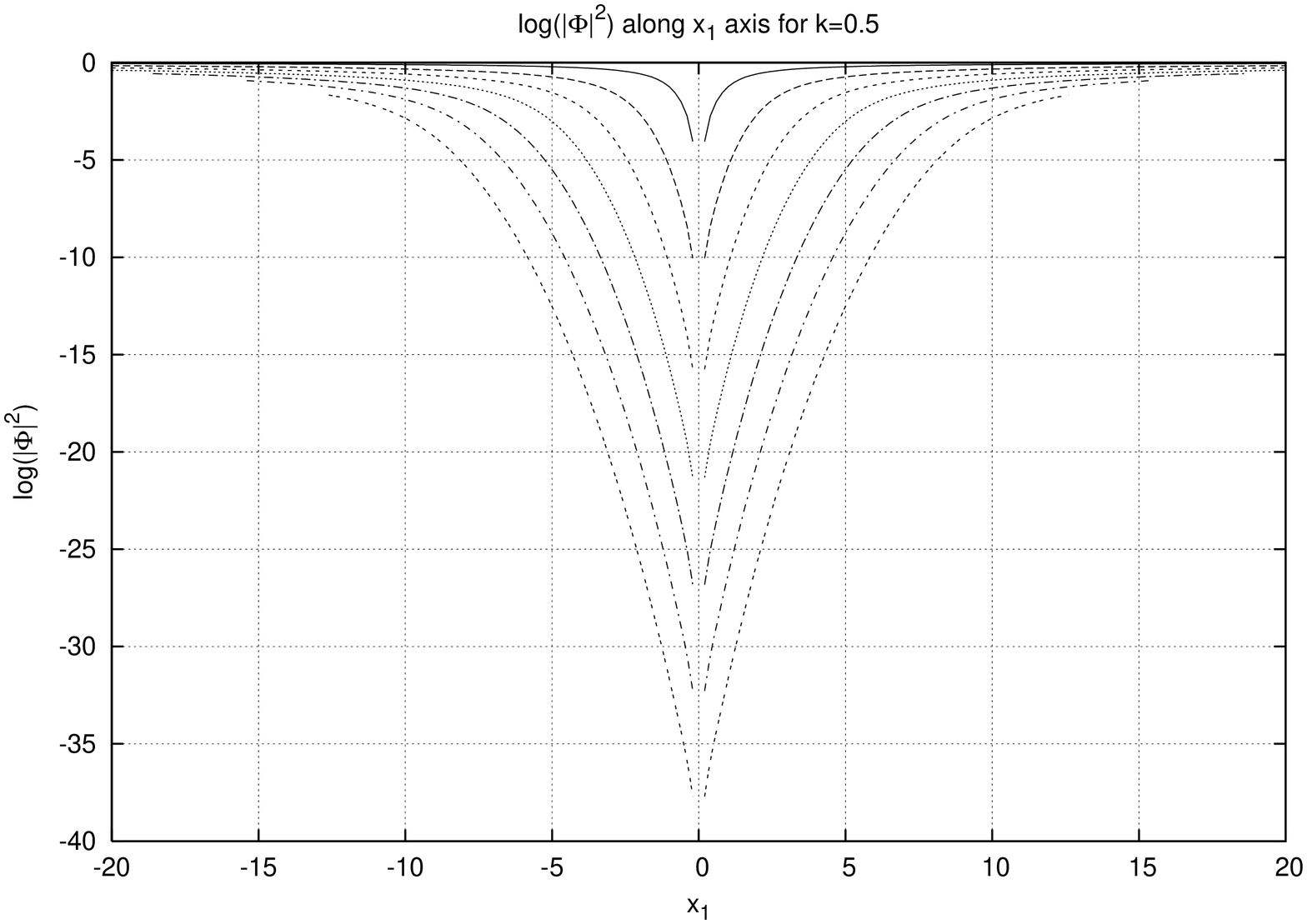}
\includegraphics[scale=0.45,angle=270]{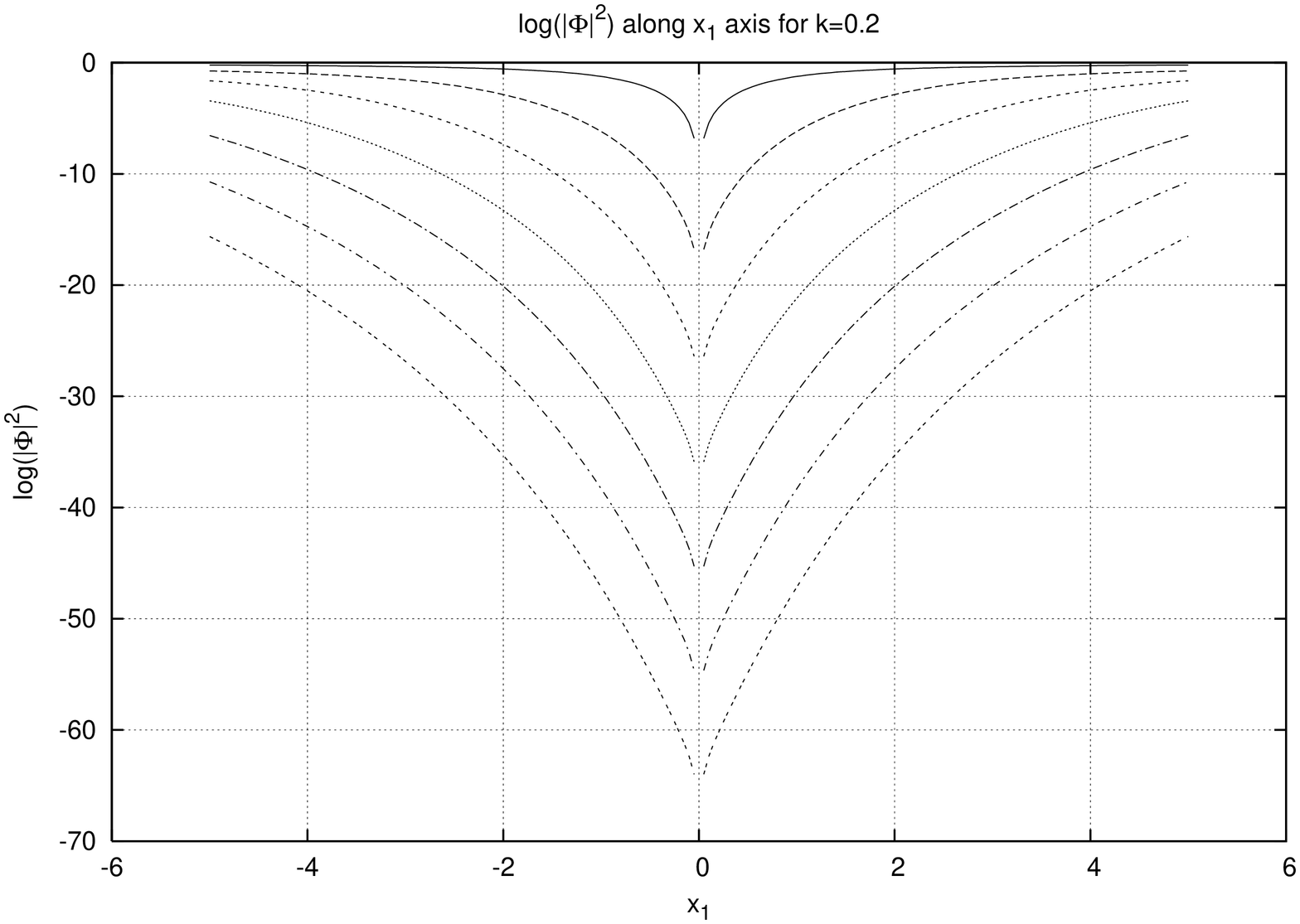}
\caption{Plots of  $\log(|\Phi|^2)$ along the $x_1$ axis for elliptic parameters $k=0.99$ (top - Fig 5a), $k=0.5$ (middle - Fig 5b) and $k=0.2$ (bottom - Fig 5c). Unlike the corresponding plots in Figure 3 along the $x_2$ axis, there is no sign of any periodic structure in the $x_1$ direction.}
\label{figure5}
\end{figure}

\begin{figure}[ht]
\vskip -2cm
\includegraphics[scale=0.45,angle=270]{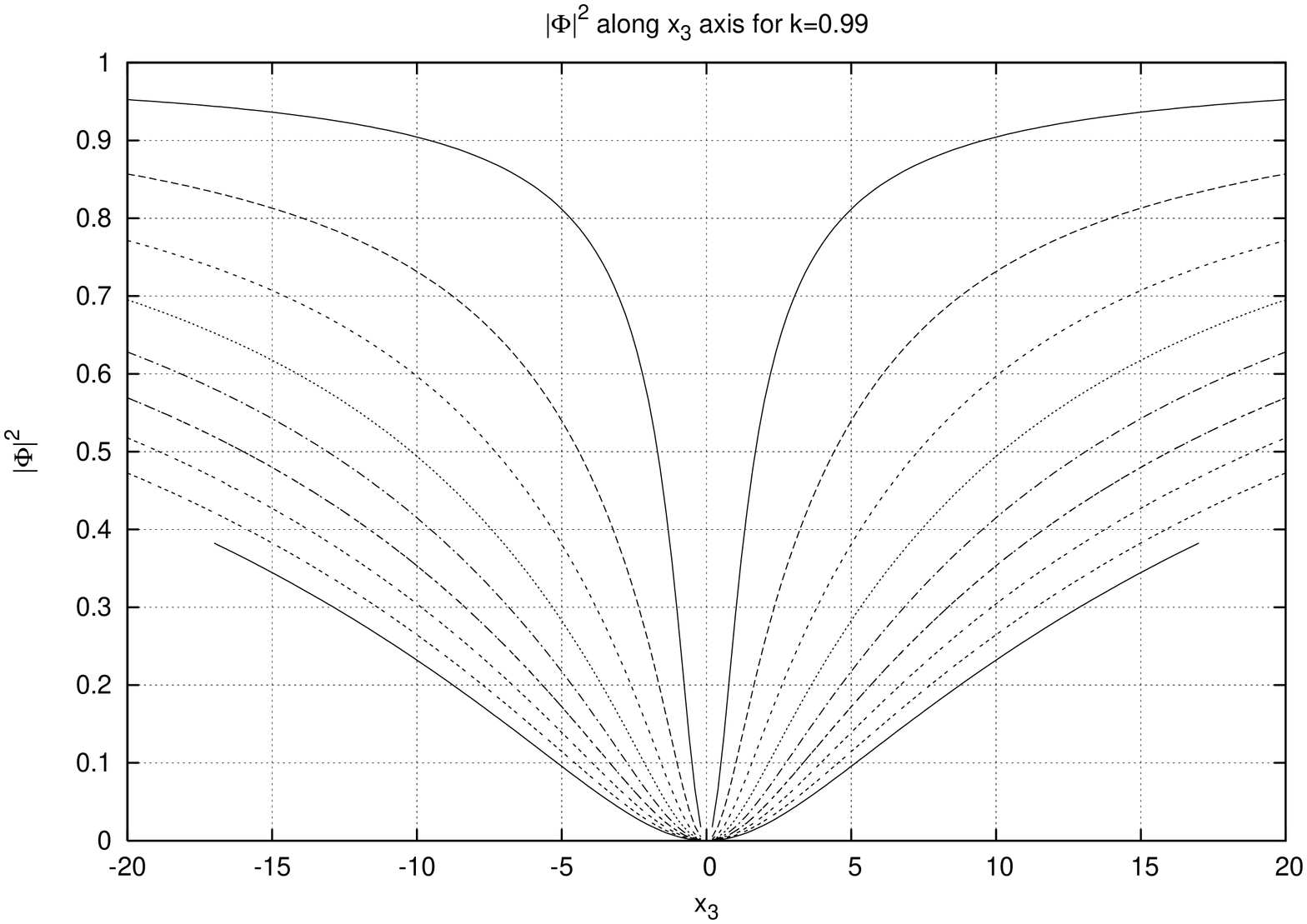}
\includegraphics[scale=0.45,angle=270]{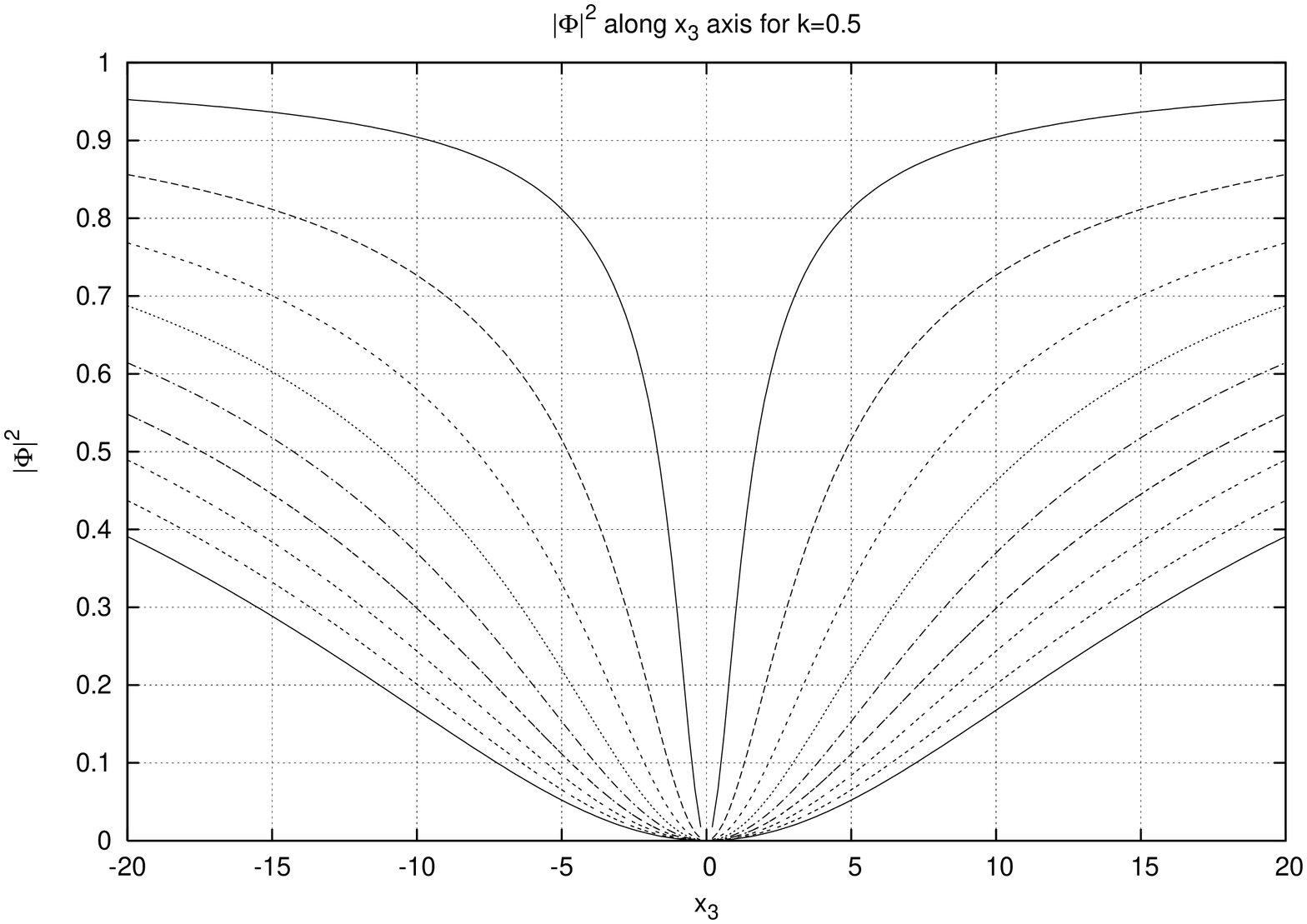}
\includegraphics[scale=0.45,angle=270]{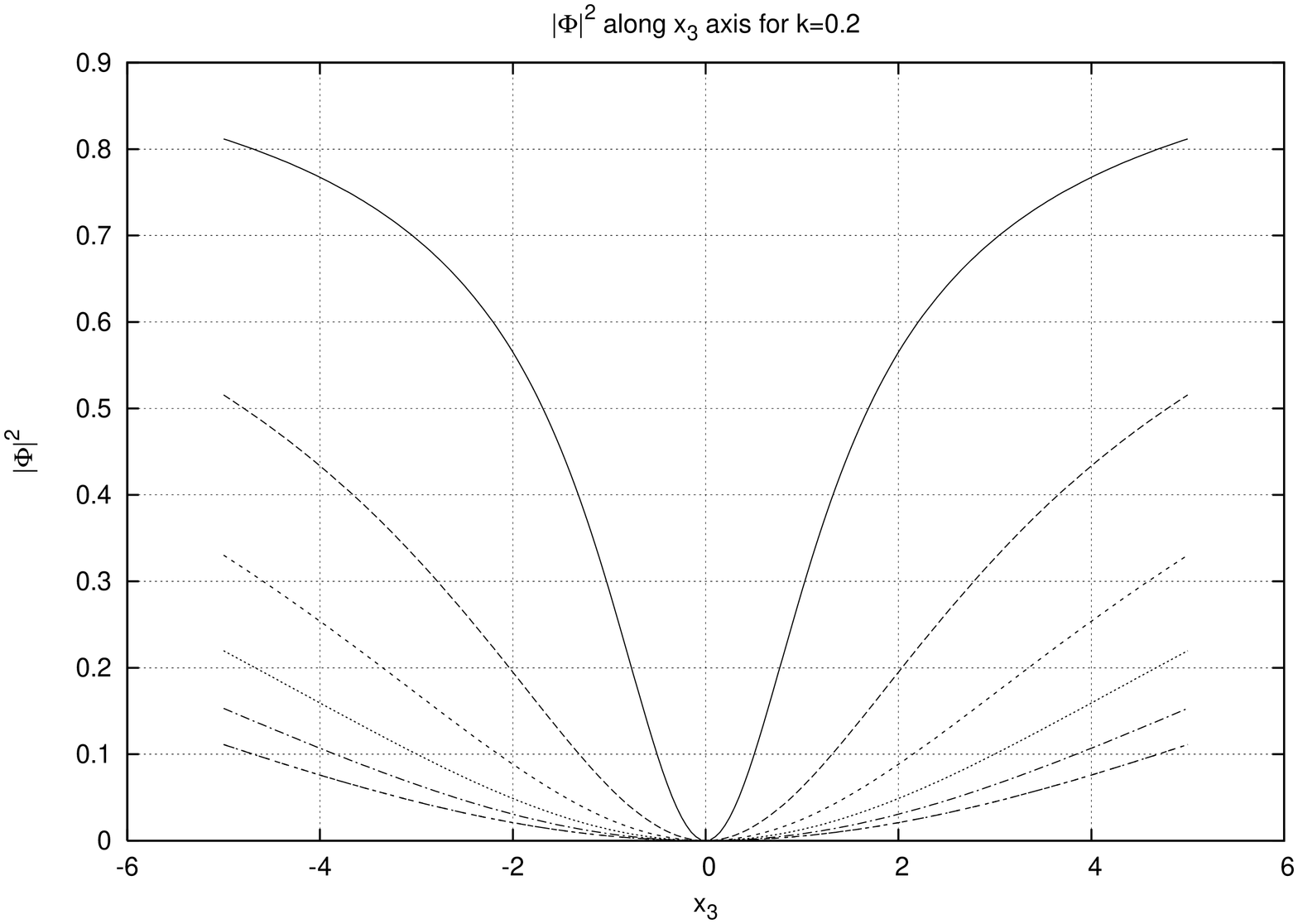}
\caption{Plots of $|\Phi|^2$ along the $x_3$ axis for elliptic parameter $k=0.99$ (top - Fig 6a), $k=0.5$ (middle - Fig 6b) and $k=0.2$ (bottom - Fig 6c). The different curves within each plot correspond to increasing $N$, starting with  $N=1$ at the top.}
\label{figure6}
\end{figure}

First, consider the behavior of $|\Phi|^2$ along the direction of the monopole chain, which for the choice of Nahm data in (\ref{nahmdata}) is the $x_2$ direction. (A simple relabeling rotates the monopole in $\vec{x}$ space). Figure 1 shows $|\Phi |^2$ along the $x_2$ axis, for various values of the monopole number, and for $k=0.99$, $k=0.5$ and $k=0.2$, respectively. 
For the $k=0.99$ case one can clearly see the array of equally spaced zeros along the $x_2$ axis. Note that the peak values of $|\Phi|^2$ between adjacent zeros is very small, and also decreases as $N$ increases. As $N$ gets large $|\Phi|^2$ vanishes within a region of length $N\beta$, outside of which it reverts to its $N$-monopole form (compare with figure 1). But for $k=0.5$ and $k=0.2$ it looks as though $|\Phi|^2$ vanishes inside this region, $-J \beta \leq x_2\leq J \beta$, already for $J=2$. This is not the case, however, as can be seen from Figure 3, which plots the logarithm of $|\Phi|^2$ along the $x_2$ axis. From these we can clearly see the periodic structure, with another two monopoles appearing on the $x_2$ axis each time $N$ is increased by 2. Notice how extremely small the peak values of $|\Phi|^2$ become between adjacent monopoles as N increases. This is an indication that there are strong core interactions, preventing $|\Phi|^2$ from getting anywhere near its asymptotic value between adjacent monopoles. We deduce the period to be  $\beta\approx 3.25$ for $k=0.99$, $\beta\approx 0.8$ for $k=0.5$, and $\beta\approx 0.3$ for $k=0.2$. This period is measured in units of the single-monopole core size, so $k=0.99$ corresponds to separated monopoles, which in the 3D Georgi-Glashow  language corresponds to the low temperature limit. For $k=0.5$ the separation and core size are comparable, while for $k=0.2$ the separation is less than (roughly one third) the core size. In the notation of Ward's recent paper \cite{ward}, the cases $k=0.99$, $k=0.5$ and $k=0.2$ correspond to small, medium and large values of the parameter $C$ defining the ratio of the single-monopole core size to the separation.

The plots in Figure 2 should be contrasted with the behavior along the $x_1$ axis, shown in Figure 4. For $k=0.5$ and $k=0.2$ the plots are very similar to those along the $x_2$ axis, with $|\Phi|^2$ vanishing inside a region of size of order $N\beta$. But for $k=0.99$ the $x_1$ axis behavior is very different from that along the $x_2$ axis. This is the limit where the 
individual monopoles are separated by a distance greater than the core size of a single monopole. To be sure there is not some small scale structure hidden in these plots, in Figure 5 we show $\log(|\Phi|^2)$ along the $x_1$ axis. In contrast to the plots in Figure 3, there is clearly no small scale structure along the $x_1$ direction. This also serves as a check that the extremely fine structure along the $x_2$ axis is genuine, rather than a numerical artifact. 

The dependence along the $x_3$ axis is shown in Figure 6. One might expect that in the large N limit the monopole solution within the central $x_2$ period would become radially symmetric in the transverse $(x_1, x_3)$ plane. However, this is not necessarily the case. For $k=0.99$ the dependence along the $x_1$ and $x_3$ axes is similar, as can be seen by comparing the first plot in each of Figures 4 and 6, although there is somewhat flatter behavior near the origin along the $x_1$ axis. But for $k=0.5$ and $k=0.2$ there is a distinct asymmetry in this transverse plane. The asymmetry is more pronounced as $N$ increases, and is also more pronounced for smaller values of $k$. Smaller $k$ means smaller period so there is greater overlap, which presumably explains the increased asymmetry. 
Indeed, for $k=0$ the solution becomes axially symmetric in the $(x_1, x_2)$ plane (with our convention of axes) \cite{prasad,manton}. This can be seen in Figs. 2c and 4c, which plot $|\Phi|^2$ for $k=0.2$ along the $x_2$ and $x_1$ axes, respectively, and which are indistinguishable from one another on this scale, while being very different from the $x_3$ behaviour in Fig. 6c. This  should be contrasted with the $k=0.99$ and $k=0.5$ cases shown in Figs. 2a,b and Figs 4a,b, where there is a distinct difference between the behaviour of $|\Phi|^2$ along the $x_2$ and $x_1$ directions.

Although all these plots are along the axes, we found them to be representative of what happens within a cubic region of size $\sim (N \beta)^3$ centred on the origin. Thus, we make the following empirical observations:

1. In the large $N$ limit, $|\Phi|^2$ vanishes along the $x_1$ and $x_2$ axes within some region of order $(N \beta)^2$ centred on the origin. This also implies that part of the action/energy density,  $\left(\partial_1^2+\partial_2^2\right) |\Phi|^2$, vanishes within this region. 

2. In the large $N$ limit, $|\Phi|^2$ exhibits three different behaviors along the $x_3$ axis. At short distance the behavior is quadratic in $x_3$, with a coefficient that decreases with $N$ like $\frac{1}{N^{3/2}}$. At very long distances $|\Phi | \sim  1- \frac{N}{|x_3|} $, as expected outside the core of an $N$-monopole. But in the intermediate region it is not yet clear what the dependence is. It can be fit well in this regime by a linear behavior or by a logarithm. 

3. For period $\beta$ much greater than 1 (the single-monopole core size) there is some remnant of symmetry in the transverse $(x_1, x_3)$ plane, but this symmetry is badly broken for $\beta<1$.

Could these structures, which are clearly equally spaced in the $x_2$ direction for any finite $N$, survive as finite action/energy periodic structures in the infinite $N$ limit? These numerical observations show that there is considerable spreading of the core region in the transverse $(x_1, x_3)$ plane. Moreover, for $\beta<1$ this appears to be almost one-dimensional. In this case, it might be possible to obtain a finite action/energy in the large $N$ limit. However, this would be at the price of the solution ``washing itself out'', because as $\nabla^2 |\Phi|^2$ spreads further out along the $x_3$ direction with increasing $N$, correspondingly its peak value decreases. Thus, while this solution might appear to have a finite action/energy in the large $N$ limit, the action/energy density delocalizes and shrinks away to nothing. 

We reached a limit of simple numerical analysis in obtaining the plots reported here. It would be interesting now to investigate these empirical hints analytically at large $N$, possibly using a semiclassical large spin $J$ approach, and also by focussing directly on the $\beta\gg 1$ and $\beta \ll 1$ limits.

\section*{Acknowledgments}

We thank Alex Kovner and Kimyeong Lee for discussions, and we acknowledge the support of the US DOE through grant 
DE-FG02-92ER40716. We are very grateful to Paul Sutcliffe for helpful comments and correspondence.

\end{document}